\documentclass[10pt, oneside, letterpaper]{article}

\usepackage{mathrsfs}
\usepackage[utf8]{inputenc}
\usepackage{graphicx}
\usepackage[usenames, dvipsnames]{xcolor}
\usepackage[letterpaper,
            left=1in,
            right=1in,
            top=1in,
            bottom=1in]{geometry}
\usepackage{setspace}
\usepackage{textcomp}
\usepackage[symbol]{footmisc}
\usepackage{hyperref}
\usepackage{amssymb}
\usepackage{amsmath}
\usepackage[section]{placeins}
\usepackage{xcolor}
\usepackage[export]{adjustbox}
\usepackage{setspace}
\usepackage{natbib}
\usepackage{sectsty}
\usepackage{titlesec}
\usepackage[]{algorithm2e}
\usepackage{fancyhdr}
\usepackage{tabularx}
\usepackage{multirow}
\usepackage{xurl}

\renewcommand{\abstractname}{ABSTRACT}
\renewenvironment{abstract}
 {\small
  \begin{center}
  \bfseries \abstractname\vspace{-.5em}\vspace{0pt}
  \end{center}
  \list{}{
    \setlength{\leftmargin}{0.5in}%
    \setlength{\rightmargin}{\leftmargin}%
  }%
  \item\relax}
 {\endlist}

\begin{document}

\titleformat*{\section}{\large\bfseries}
\titleformat*{\subsection}{\bfseries}
\titleformat*{\subsubsection}{\itshape}

\onehalfspacing
\begin{center}
\large{\textbf{Crime reduction through public healthcare: Interpretable machine learning for mental health service impacts in Greater London}} 
\end{center}

\bigskip

\begin{center}
\large{Nadine F{\"a}ssler$^{\mathrm{1}}$, Ben Moews$^{\mathrm{1,2,3}}$}

\bigskip

\begin{small}
{$^{\mathrm{1}}$Business School, University of Edinburgh, United Kingdom}\\
{$^{\mathrm{2}}$Centre for Statistics, University of Edinburgh, United Kingdom}\\
{$^{\mathrm{3}}$Scottish Centre for Crime and Justice Research, United Kingdom}\\
\end{small}

\end{center}

\bigskip

\doublespacing

\begin{abstract}
{The relationship between crime, mental health service access, and socioeconomic deprivation in publicly-funded healthcare systems allowing impactful policy interventions offers an alternative lens to crime prevention that remains underexplored. We address this critical gap through an analysis of street-level crime data, mental health referral information, and socioeconomic metrics across Greater London, using both traditional statistical methods and machine learning techniques to identify relevant relationships and spatial patterns to reveal a persistent positive association between crime rates and mental health referrals as a proxy for service access. The prevailing prevention hypothesis is contrasted with a nuanced U-shaped relationship suggesting a contrast between preventive effects at lower service levels and demand-driven responses to crime exposure for higher referral rates. Subsequent analyses, focussing on explainable artificial intelligence, show distinct crime category patterns, with a cluster analysis identifying four borough typologies with distinct combinations of crime rates, mental health service access, and deprivation levels, requiring multifaceted approaches rather than universal solutions. This research provides one of the first comprehensive studies on this topic for the UK's publicly-funded healthcare system and introduces interpretation-oriented approaches to uncover the patterns essential to evidence-based policies.}
\end{abstract}

\bigskip

\begin{footnotesize}


\end{footnotesize}

\bigskip


\section{Introduction}
\label{sec:introduction}

In recent years, London's rising crime rate has frequently been featured in both media and research \citep{Wu2023}. Much of this rise has been driven by violent crime, with crime rates reaching 30.1 offences per 1,000 population in 2023/24. Given the social, economic, and public health costs of crime, particularly violent crime, developing effective prevention strategies has become a policy priority \citep{miller1996}. In response, the London Mayor's Office for Policing and Crime developed the Police and Crime Plan 2025--2029, with a strong focus on reducing violent crime through a public health framework.\footnote{\url{https://www.london.gov.uk/programmes-strategies/mayors-office-policing-and-crime-mopac}}

Crime prevention through public health improvements is based on extensive evidence of underlying determinants, demonstrating socioeconomic deprivation as an important shared aspect \citep{moore1995}. Low socioeconomic status (SES) and poverty in particular are associated with higher rates of mental health problems due to associated pressures, while also increasing the likelihood of criminal behaviour robustly across violent crime types \citep{adler1994, galloway2010, kelly2000}. Correspondingly, a successful public health approach targets prevention through interventions in the former, simultaneously improving health outcomes and reducing crime.

Based on this approach, London is increasing access to therapy and specialised models of care .\footnote{\url{https://www.london.gov.uk/programmes-strategies/communities-and-social-justice/londons-violence-reduction-unit-vru}} However, to ensure effectiveness and resource efficiency, it is crucial to understand underlying relationships in a localised context. While early research on institutionalised populations consistently shows an association between mental health and crime, with three times higher arrest rates than the general population, research on community-based samples yields mixed findings \citep{link1992, silver2008, halle2020}.

As a result, research examining the effect of population-level access to these services is focussed on identifying causal effects through largely US-based natural experiments and quasi-experimental designs, such as variations in Medicaid eligibility or expansions, demonstrating mental health coverage loss increasing the probability of incarceration, and vice-versa \citep{wagner2020}. Related studies, for example on office-based mental healthcare and, so far, limited applications of machine learning in the literature, equally show relevant relationships \citep{deza2022b, ramezani2022}.

Several critical research gaps remain. As these studies primarily focus on the US healthcare system with its privatised structure, the relationships in systems like the National Health Service (NHS) in the UK remain largely unexplored, and structural differences might affect the applicability. Existing research also primarily relies on traditional statistical approaches \citep[see, for example,][]{andresen2006, murray2001, suss2023}, with a lack of machine learning techniques to capture complex non-linear relationships.

This study addresses these gaps by examining the relationship between mental health service access and crime rates within a universal healthcare system, using the Greater London Area as a densely populated urban centre with large-scale data availability. To develop informed policy recommendations, we further examine how the relationship between mental health service access and crime varies across different crime categories, as well as the role of shared socioeconomic and demographic factors. Finally, by identifying distinct geographical typologies, our work provides insights for tailored prevention strategies.

Employing both statistical methods and interpretable machine learning, we first establish the examined relationship through a cross-sectional analysis and gain further insights through predictive validity and a cluster analysis for geospatial patterns. We use a combined dataset of street-level crime data from the UK Police, mental health referral data as a proxy for mental health service access provided by NHS England, and socioeconomic indicators from the Ministry of Housing, Communities and Local Government.

This work provides one of the first comprehensive studies of this relationship within the UK's context, expanding the literature beyond its US focus, and offers several original contributions. Uncovering complex relationships across 32 London boroughs, our results challenge prevailing assumptions, with a U-shaped relationship involving supply and demand-driven responses. We also observe and interpret increased predictive model complexity for violent versus property crime through the lens of existing theory. It further contributes to the adaption of both supervised and unsupervised machine learning techniques to preserve interpretability over pure predictive power as an alternative to established approaches, and offers policy-makers practical tools for developing targeted, evidence-based crime prevention strategies.


\section{Literature Review}
\label{sec:litreview}

\subsection{Public health approaches to crime prevention} 
\label{subsec:2.1}

The criminal justice system in the UK during the 1960s was built on rehabilitation, with the consensus supporting penal welfarism as a rehabilitation-focussed approach treating underlying causes \citep{garland2004}. However, by the mid-1970s, this consensus began to decline amid increasing scepticism about its effectiveness. \citet{becker1968} introduces the rational choice theory of crime as an alternative perspective, reframing criminal offence as a calculated decision weighing the associated costs and benefits. 

This model implies that crime can be addressed by increasing either the associated risk through punishment or the associated opportunity cost through, for example, improved employment and economic opportunities. This allows for both punitive and preventative strategies, but policy-makers largely reduced the model to its punitive component, leading to a steep rise in the English and Welsh prison populations, particularly through longer sentences for violent offences and persisting into the 21st century despite declining conviction rates \citep{garland2004, newburn2007}.

Other countries have continued to adopt a rehabilitative approach, with the Norwegian penal system often being cited for its focus on restorative justice, humane prison conditions, and community reintegration, resulting in some of the lowest global recidivism rates \citep{pratt2008}. There is increasing evidence for the limitations of punitive approaches, spanning from social costs and systemic strains on the justice system to disproportionate impacts on marginalised communities \citep{cullen2011, carr2017}.

This has prompted a reevaluation of \citet{becker1968}'s approach, with scholars and practitioners advocating for strategies addressing the complex social, economic, and psychological determinants of criminal behaviour \citep{cullen2016}. This shift is reflected in the increasingly adopted view of criminal behaviour as a public health issue, which reframes crime not as an individual choice or moral failure, but as a response to broader societal and health inequalities that requires investments in mental health and community-based support \citep{moore1995, bucerius2021}.

SES and poverty are important social determinants of both health and criminal behaviour, including restricted access to housing, healthcare, and education and employment opportunities, as well as higher crime exposure \citep{finegan2020}. This leads to higher prevalences of obesity, cardiovascular disease, and mental health disorders \citep[see][]{adler1994, franks2011, muntaner2004}. \citet{silva2016} conduct a review of 150 studies to examine associations between both individual and area-level socioeconomic factors and mental health outcomes, finding a consistent negative effect.

Socioeconomic deprivation has also been associated with access to, and outcomes of, psychological treatment. UK-based studies using the English Index of Multiple Deprivation have found that individuals living in socioeconomically deprived neighbourhoods not only have a lower probability of accessing mental healthcare, but when they do access care, their likelihood of recovery is reduced \citep{delgadillo2018, clark2017}. \citet{finegan2020} demonstrate this two-fold effect analysing healthcare records from 44,805 UK patients, showing that neighbourhood income and crime rates are significantly correlated with poorer treatment outcomes, even after controlling for individual patient-level variables. 

Similarly, income inequality has been shown to affect health outcomes \citep{wilkinson2007, kondo2009}. This extends to mental health through stressors such as feelings of relative deprivation and reduced sense of control, which in turn have been linked to adverse mental health outcomes, with a survey analysis of 12 developed countries finding a three times higher prevalence of mental health conditions with higher income inequality \citep{pickett2010}.

These determinants also affect criminal behaviour, with studies demonstrating how low SES and poverty during childhood increase the likelihood of later substance abuse and criminal behaviour \citep{galloway2010, fergusson2004}. An increased probability of criminal behaviour also links this to higher rates of victimisation, creating a cycle affecting community-level violence, and the relationship between poverty and criminal offences remains robust across different violent crime types \citep{berg2012, kelly2000}. Strategies targeting these risk factors could have significant benefits for both mental health and criminal behaviour. However, shared social determinants do not rule out additional direct causal pathways, with mental health status itself independently affecting criminal behaviour through psychological or behavioural mechanisms. Understanding both pathways is crucial for informed, evidence-based policymaking.

\subsection{Empirical evidence on a direct relationship}
\label{subsec:2.2}

Empirical research has examined potential direct influences of mental health on criminal behaviour, beyond shared social determinants establishing one pathway. If mental health problems lead to increased criminal activity through psychological or behavioural mechanisms such as reduced impulse control or coping strategies, this relationship should persist when controlling for shared determinants. The corresponding body of literature has evolved considerably over recent decades, while research prior to the 1980's typically finds little evidence of a systematic link \citep{link1992}.

Early evidence to the contrary stems from studies focussing on criminal behaviour and arrest rates of institutionalised individuals, finding that persons with a history of psychiatric hospitalisation are more likely to have been convicted of a crime and three times more likely to be arrested \citep{hodgins1996, link1992}. These insights are replicated by other works, for example by \citet{nao2017}, showing that 37\% of prison populations in England and Wales report emotional and mental health issues. Cross-national research also suggests a factor of more than three with regard to the detainment of individuals with mental health disorders in prisons compared to mental health institutions \citep{fazel2016}.

Prior research establishes specific associations; \citet{vinkers2011} examine defendants with mental health disorders and find crime type associations are most pronounced for arson, battery, homicidal attempts and threats, and both sexual and violent crimes. Similarly, \citet{ponde2014} identify relationships between specific diagnoses and crime types within prison populations, which is furthered by a cross-sectional study investigating data from the Survey of Inmates in State and Federal Correctional Facilities covering 17,248 inmates across 275 state and 40 US federal prisons \citep{silver2008}. The latter incorporates controls for prior offences, substance use, demographics, and victim characteristics, finding associations between a history of mental health problems and assaultive violence and sexual offences compared to property, drug, and other crimes, supporting a deviance hypothesis with regard to anti-normative crimes.

The generalisability of these findings is impacted by a focus on high-risk institutionalised populations, introducing selection bias towards more serious and chronic offenders. A small number of studies have addressed this through a more representative community-based approach, for example \citet{corrigan2005} on a 3--4 times increased likelihood of engagement in violent behaviour for individuals with mental health diagnoses, confirming earlier research on a four-fold increase in violent behaviour observed by \citet{swanson1990}. Measures of violences in both cases are, however, broadly defined, thus potentially inflating the observed relationship through the inclusion of, for example, encounters with the police involving physical altercations requiring medical attention, criteria that may include acts of self-defence.

These results are contrasted by more recent research comparing individuals with and without mental health disorders, revealing no significant differences in arrest or conviction rates, as well as no significant associations between specific disorders and crime types. Overall violent crime shows an engagement of 21.1\% versus 7.8\% in the control group, although the sample size of $N = 121$ (17 for violent offences)  and reliance on self-report data provide a limitation \citep{halle2020}. 

While mixed findings in the literature highlight the complexity of this topic, the majority of studies indicate some form of relationship between mental health and criminal behaviour, albeit more nuanced than initially suggested by institutional studies. The difficulty of isolating direct causal effects from confounding factors, meaning shared underlying determinants such as poverty and neighbourhood deprivation, provides a challenge that affects the validity of inferences when not properly accounted for. Notably, most studies also focus on the United States, where healthcare access is governed by insurance-based models, raising concerns about the generalisability of these findings to universal healthcare systems. 

While the exact nature of the relationship between mental health and criminal behaviour remains somewhat inconclusive, both theoretical pathways, the correlation through shared determinants and direct effects, suggest that mental health interventions might influence criminal behaviour. This has led researchers and policymakers to examine whether mental health interventions can reduce criminal activity.

\subsection{Population-level access to mental health services}
\label{subsec:2.3}

The relationship between population-level access to health services and criminal behaviour when controlling for shared social determinants has, more recently, been studied through natural experiments and quasi-experimental designs. For example, \citet{wagner2020} investigate Medicaid expansions as a natural experiment, using an event study approach with state and year-fixed effects and variations in the timing of expansions targeting populations with high rates of mental illness in 1995--2010. 

The analysis demonstrates a statistically significant per-capita crime incident reduction of 1.7\% in the year of the Medicaid expansion, and 4.25\% the following year. This effect is measured to be strongest for property crimes, with a smaller effect for violent crimes, which is an expected outcome due to property crimes accounting for 88\% of crimes in the corresponding data. There is, however, supporting evidence from other studies focussing on substance abuse instead of general mental health services, suggesting that Medicaid coverage leads to a reduction in criminal activity \citep[see, for example,][]{wen2017}.

Exploring the long-term implications of healthcare access, \citet{hendrix2022} equally focus on Medicaid to investigate the relationship between childhood coverage and early-adulthood crime rates through a quasi-experimental design. Based on variations in Medicaid expansions across states and birth cohorts, the authors report that an additional year of childhood Medicaid eligibility results in a 7\% reduction in property crime for the ages 19--24, although no statistically significant effect is shown for violent crimes. From an economic point of view, the study also shows that there is a \$0.19 return per US Dollar spent on childhood Medicaid eligibility, using social cost of crime estimates by \citet{miller2021}.

The cost of crime generally encompasses three types of cost; These are (i) victim-related costs directly caused by criminal behaviour such as property damage, medical or mental health treatment, and productivity losses; (ii) societal response costs including policing, courts, and victim services; and (iii) costs incurred by offenders such as lost wages \citep{cohen2020}. Therefore, the monetary benefit to society from reduced crimes extends far beyond direct incarceration savings.

Another approach in the literature is the study of geographic availability of mental health services, using the number of office-based mental health providers as a proxy. \citet{deza2022b} implement a two-way fixed effects regression model with US county-level panel data from 1999--2014 to examine how variations in availability affect the rates of Part I crimes covering violent and property crimes, revealing that an increase of 10 office-based mental healthcare providers results in 1.6 fewer incidents per 10,000 people. Accounting for crime type-specific social costs, this translates to a 2.2\% reduction in cost-adjusted crime per capita. Using the same approach on juvenile populations, \citet{deza2022a} find that 10 additional mental health offices per county reduce the social costs of juvenile arrests by 2.3--2.6\%.

While most population-level studies apply traditional statistics, a small body of research employing machine learning approaches has recently emerged to leverage advantages in identifying complex patterns. In this context, to study the relationships of mental health service availability, community characteristics, demographic factors, and incarceration rates across all 3,141 US counties, \citet{ramezani2022} implement a combination of techniques after variable selection through a multidisciplinary expert panel. In preparation for a beta regression, dimensionality reduction is first achieved with the random forest algorithm, while L1 regularisation is used to select a final set of variables.

These experiments show mental healthcare factors as significant predictors of prison populations, with a 4\% reduction of the jailed-to-non-jailed population ratio for one-unit increases in psychiatrists per capita. The authors also find that counties with lower mental health service accessibility are associated with higher incarceration rates. Population-level studies such as this provide evidence for the effect of mental health service accessibility on crime prevention. However, the majority of the literature relies on traditional statistical approaches and could overlook complex and non-linear relationships.

At the same time, the black-box nature of many machine learning algorithms can be a hindrance for inference, motivating a focus on explainable artificial intelligence through simpler models. Just as importantly, the stark focus on US data in the literature, due to the different healthcare system when compared to the UK, raises questions about the applicability in publicly funded systems and, consequently, the effectiveness of policy implications for crime prevention.


\section{Methodology}
\label{sec:methodology}

\subsection{Cross-Sectional Analysis}
\label{subsec:3.1}

We employ several approaches, spanning both statistics and machine learning, in this work. We first perform a bivariate ordinary least squares (OLS) regression to estimate the direct association between the dependent and independent variables. The baseline model is specified as
\begin{equation}
    c_i = \beta_0 + \beta_1 h_i + \varepsilon_i,
\end{equation}
where $c_i$, $h_i$ and $\varepsilon_i$ represent the crime rate, mental health service access proxied by therapy referral rates, and the error term, for a given borough $i$. The intercept is $\beta_0$, while $\beta_1$ is the slope coefficient indicating the change in crime rates associated with a one-unit increase in referrals \citep{james2021}. To assess whether this relationship persists when accounting for shared social determinants, we extend this as
\begin{equation}
c_i = \beta_0 + \beta_1 h_i + \beta_2 s_i + \beta_3 d_i + \beta_4 m_i + \varepsilon_i.
\end{equation}
The variables $s_i$, $d_i$, and $m_i$ serve as controls for borough-level deprivation, youth demographics, and ethnic composition. Holding other variables constant, $\beta_1$ denotes the partial effect of mental health referrals on crime rates. We estimate coefficients using OLS, with the objective to minimise the sum of squared residuals to obtain unbiased parameter estimates \citep[see][]{james2021},
\begin{equation}
\min_{\beta_0, \beta_1, \beta_2, \beta_3, \beta_4} \sum_{i=1}^{n} \left( c_i - \beta_0 - \beta_1 h_i - \beta_2 s_i - \beta_3 d_i - \beta_4 m_i \right)^2.
\end{equation}
There are several assumptions; (i) linearity between dependent and independent variables, (ii) normality of the residuals, (iii) homoscedasticity, and (iv) independence of the residuals from each other \citep[see][]{heumann2016}, which we evaluate via residual plots, the Shapiro-Wilk test and Q-Q plots, Levene's test, and Moran's $I$ \citep{heumann2016, shapiro1965}. The Shapiro-Wilk test uses
\begin{equation}
    W=\frac{(\sum_{i=1}^n a_ix_{(i)})^2}{\sum_{i=1}^n(x_i-\bar x)^2},
\end{equation}
where $x_{(i)}$ is the $i$-th order statistic, $\bar x$ the sample mean, and $a_i$ the Shapiro-Wilk coefficients. For Levene's test, to test the null hypothesis $\sigma_1^2=\sigma_2^2=\dots =\sigma_n^2$, we define subsets of boroughs with similar predicted crime rates by dividing the regression model's fitted values into quartiles. After calculating the absolute difference between each value and its median, we apply a one-way analysis of variance (ANOVA) via
\begin{equation}
    W=\frac{(N-k)}{(k-1)}\cdot \frac{\sum_{i=1}^n n_i(\bar{Z_{i}}-\bar{Z})^2}{\sum_{i=1}^k \sum_{j=1}^{n_i}(Z_{ij}-\bar{Z_{i}})^2}.
\end{equation}
Here, $n_i$ is the sample size in group $i$ for \textit{k} groups, $N$ the total sample size,  $Z_{ij}=|Y_{ij}-\bar Y_i|$ the absolute deviation of observation $j$ in group $i$ from the group median $\bar Y_i$, $\bar{Z_{i}}$ the mean of $Z_{ij}$ in group $i$, and $\bar{Z}$ the overall mean of $Z_{ij}$. We calculate Moran's $I$ using Queen contiguity weights to examine spatial autocorrelation, measuring the degree of spatial clustering as
\begin{equation*}
    I=\frac{n}{S_0}\cdot \frac{\sum_{i=1}^n \sum_{j=1}^n w_{ij}z_iz_j}{\sum_{i=1}^n z_i^2},
\end{equation*}
where $n$ is the number of observations, $S_0=\sum_{i=1}^n\sum_{j=1}^n w_{ij}$ the sum of spatial weights for respective boroughs $i$ and $j$ (equal to 1 if adjacent, 0 otherwise), and $z_i$ and $z_j$ the deviations from the mean residual \citep{moran1950}. We use a significance threshold of $\alpha = 0.05$ throughout these tests. Given the violation of key assumptions for multiple linear regression, we subsequently perform correlation analyses using three approaches. The Pearson correlation coefficient $r$ measures linear relationships via
\begin{equation}
    r = \frac{\sum_{i=1}^{n} (x_i - \bar{x})(y_i - \bar{y})}{\sqrt{\sum_{i=1}^{n} (x_i - \bar{x})^2} \sqrt{\sum_{i=1}^{n} (y_i - \bar{y})^2}},
\end{equation}
with $n$ as the sample size, $x_i$, $y_i$ as boroughs, and $\bar{x}$, $\bar{y}$ as the sample means. Spearman's rho, in contrast, measures monotonic relationships through value ranks instead of the values themselves, with the coefficient being calculated, with $d_i$ as the difference between the ranks of two observations, as
\begin{equation}
    \rho = 1 - \frac{6\sum_{i=1}^{n} d_i^2}{n(n^2-1)}.
\end{equation}
Finally, Kendall's tau measures the ordinal association between two observations, using $C$ and $D$ as the number of concordant and discordant pairs, respectively, to evaluate the probability of two variables having the same relative ordering \citep[see][]{field2018, heumann2016},
\begin{equation}
    \tau = \frac{C-D}{C+D}.
\end{equation}
Comparing these measures provides insights into the underlying relationship structure; for example, $\rho$ being substantially higher than $r$ suggests monotonic but non-linear patterns, highlighting the need for more flexible modelling approaches. Due to the latter, we subsequently apply polynomial regression, extending the linear framework by allowing the model to capture non-linear associations while preserving interpretability \citep{james2021}. We identify the second-degree specification for the best model fit,
\begin{equation}
   c_i = \beta_0 + \beta_1 r_i + \beta_2 r_i^2 + \beta_3 s_i + \beta_4 d_i + \beta_5 m_i + \varepsilon_i,
\end{equation}
with $r_i$ as the linear term for mental health service access, while $r_i^2$ enables the capturing of threshold effects and diminishing or accelerating returns. To examine whether different crime categories show distinct non-linear patterns, we also examine the polynomial regression separately for violent and property crime rates. We re-estimate all models using 2018 data to ensure robustness and examine first treatment rates per 100,000 people as an alternative measure to validate the primary findings using referral rates.

\subsection{Predictive Modelling}
\label{subsec:3.2}

After establishing associations within cross-sectional data, we investigate whether these relationships can be generalised across time periods, testing the stability of spatial patterns and borough-level patterns over time to inform future resource allocation decisions. Model evaluation uses the mean absolute error (MAE),
\begin{equation}
    \mathrm{MAE} = \frac{1}{n} \sum_{i = 1}^n |y_i - \hat{y}_i|,
\end{equation}
for $n$ data points, true values $y_i$, and predictions $\hat{y}_i$. As a second-degree polynomial regression model adequately captures the non-linear relationship, we first explore its use through \textit{k}-fold cross-validation. The data is randomly divided into $k$ approximately equal parts and trained on $k - 1$ parts to be tested on the remainder. In doing so, a mean squared error (MSE) is calculated for each of these ``folds'', $\{\mathrm{MSE}_1, \dots, \mathrm{MSE}_2\}$, for which the average cross-validation MSE can then be computed as
\begin{equation}
    \mathrm{MSE}_{\mathrm{CV}} = \frac{1}{k}\sum_{i=1}^{k} \mathrm{MSE}_i \ \mathrm{s.t.} \ \mathrm{MSE}_i = \frac{1}{n_i}\sum_{j=1}^{n_i}(y_j - \hat{y}_j)^2,
\end{equation}
with $n_i$ as the umber of observations in the $i$-th fold, and $y_i$ and $\hat{y}_i$ denoting the true and predicted target values, respectively \citep{james2021}. Balancing computational efficiency and reliable performance estimation, we use $k = 5$, but clear evidence of overfitting highlights the risk of strong in-sample performance, as the polynomial regression model used may not generalise well to new data.

Decision trees are well-suited due to their inherently non-linear nature, interpretability and lack of functional assumptions \citep{kuhn2013, rudin2019}. Regression using these models is based on segmenting the independent variable space into non-overlapping regions, with predictions based on mean response of training observations \citep{james2021}. The algorithm seeks to divide the $N$-dimensional feature space $\{X_1, X_2, ..., X_N\}$ into $J$ regions $R_1, R_2,...,R_J$ that minimise the residual sum of squares (RSS),
\begin{equation}
\mathrm{RSS} = \sum_{j=1}^{J} \sum_{i \in B_j} (y_i - \hat{y}_{R_j})^2,
\end{equation}
with $y_i$ as the true value of the response variable for the $i$-th observation in region $R_j$ and $\hat{y}_{R_j}$ as the mean response value in region $R_j$. Given the computational infeasibility of evaluating all possible partitions, decision trees adopt a top-down, greedy approach known as recursive binary splitting. At each node split, the algorithm identifies the predictor $X_j$ and threshold value $s$ for the highest RSS reduction when partitioning the data into subsets $\{{X|X_j<s}
\}$ and $\{X|X_j\geq s\}$. This is done recursively until a stopping criterion is met, creating a hierarchical set of decision rules as a tree structure \citep{james2021}.

Due to the model's tendency to overfit, particularly for small samples, with complex trees memorising noise, we optimise hyperparameters using a grid search with \textit{k}-fold cross-validation for maximum tree depth, minimum samples per leaf, and minimum samples to split, selecting the lowest cross-validation MSE \citep{hastie2009}. This approach effectively balances model fit and generalisation by limiting model complexity, trading a potential reduction in predictive accuracy for reduced overfitting and variance as well as interpretability. We investigate violent and property crimes separately to examine whether crime categories exhibit different pathways, and apply the same robustness tests as before.

\subsection{Clustering Analysis}
\label{subsec:4.1}

To explore heterogeneity, we extend our work through cluster analysis to identify subgroups exhibiting similar behaviour across key variables. We employ \textit{k}-means, as it offers high computational efficiency and scalability compared to hierarchical clustering and, in contrast to probabilistic or fuzzy clustering methods, provides non-overlapping cluster assignments, which facilitates interpretation and communication of the results. The algorithm makes no strict assumptions about the data distribution and partitions the dataset into a chosen number of $K$ clusters such that the total within-cluster sum of squares is minimised via
\begin{equation}
\min_{C_1, \dots, C_i} \sum_{i = 1}^{k} \sum_{j \in C_i} ||x_j - \bar{x}_i||^2 \ \mathrm{s.t.} \ \bar{x}_i = \frac{1}{|C_i|}\sum_{j \in C_i}x_{ji},
\end{equation}
where $C_i$, represents the $i$-th cluster, $x_j$ the $j$-th observation, $\bar{x}_i$ the mean vector (or centroid) of cluster $i$, and $||x_j - \bar{x}_i||$ the squared Euclidean distance between each observation $j$ and the centroid of cluster $i$ \citep{hastie2009}. As there are approximately $k^n$ possible partitions for $n$ observations and $k$ clusters, finding the globally optimal solution is computationally intractable, which $k$-means circumvents with an iterative procedure searching for a local optimum. After initiating with randomly-assigned clusters, the algorithm alternates between computing the centroid for each cluster $k$ and reassigning data points to the their closest centroid, until convergence is reached by assignments no longer changing \citep{james2021}. 

We assess cluster stability through five random seeds $\{42, 123, 456, 789, 999\}$\footnote{The first seed representing, as per specialised literature, the answer to life, the universe, and everything, while the remainder are unfortunately all-too-common exercises in secure password generation.} and evaluate consistency with the Adjusted Rand Index (ARI), where scores close to 1 or 0 indicate stable clustering or high initialisation sensitivity and unstable configurations, respectively \citep{steinley2004}. Our results suggest largely consistent clustering solutions. Their number $k$ is selected via the elbow criterion and silhouette analysis, with the former identifying the point at which increases yield diminishing returns in variance reduction and the latter maximising, with the silhouette coefficient $s(i)$ for a given observation $i$,
\begin{equation}
    \mathrm{SC}=\max_k \bar s(k) \ \mathrm{s.t.} \ s(i)=\frac{b(i)-a(i)}{\max(a(i),b(i))}.
\end{equation}
Here, $a(i)$ denotes cohesion as the average intra-cluster distance and $b(i)$ separation, meaning the average distance to the nearest neighbouring cluster, while $\bar{s}(k)$ represents the average silhouette score for $k$ clusters. Prior to analysis, as range differences would otherwise distort results, we apply $z$-score normalisation to all variables, with an estimated mean $\mu$ and estimated standard deviation $\sigma$,
\begin{equation}
    z_i=\frac{x_i-\mu}{\sigma}.
\end{equation}
The validity of solutions is then evaluated using the Kruskal-Wallis $H$ test, a non-parametric method for comparing the distributions of a continuous variable across more than two groups \citep{kruskal1952},
\begin{equation}
    H = \frac{12}{n(n+1)} \sum_{i=1}^{k} n_i \left(\varphi_i - \frac{n+1}{2}\right)^2,
\end{equation}
where $k$ is the number of clusters, $n_i$ the number of observations in cluster $i$, $n$ the total number of observations, and $\varphi_i$ the average rank of observations in cluster $i$. We also apply the same robustness checks.


\section{Analysis and Results}
\label{sec:analysis_results}  

\subsection{Data and Analytical Framework}
\label{subsec:4.2}

Focusing on the Greater London area, this study adopts a borough-level analysis, corresponding with local government administrative boundaries and policy-making structures. Our analysis spans 32 boroughs, excluding the City of London due to missing data and its distinct non-residential characteristics. We use street-level crime data from the UK Police open data portal \footnote{\url{https://data.police.uk}}, providing comprehensive monthly statistics from the Metropolitan Police Service and City of London Police for April 2017 to March 2020. 

This includes geographic coordinates, Lower Layer Super Output Area (LSOA) identifiers, crime type, and outcome category. Our primary dependent variable is the crime rate per 100,000 population, including violence and sexual offences, robbery, burglary, theft from person, vehicle crime, bicycle theft, shoplifting, and criminal damage and arson, aggregatable into violent and property crimes \citep{deza2022b}.

We match crime incidents to boroughs using LSOA identifiers, with 26,563 (1.45\%) of 1,834,748 entries missing geographic information. Given a random distribution of missing LSOA identifiers, we exclude these without introducing systematic bias. Finally, crime incidents are aggregated annually and standardised using Office for National Statistics (ONS) mid-year population estimates.\footnote{\url{https://www.ons.gov.uk/peoplepopulationandcommunity/populationandmigration/populationestimates}}

The standardised mental health referral rate per 100,000 population, obtained from NHS England's Improving Access to Psychological Therapies (IAPT) programme\footnote{\url{https://digital.nhs.uk/data-and-information/publications/statistical/nhs-talking-therapies-for-anxiety-and-depression-annual-reports}}, is standardised and used as a proxy due to the lack of direct measures. This includes annual referral counts for April 2017 to March 2020, reported at Clinical Commissioning Group (CCG) level for 32 CCGs coinciding with London boroughs.

Our analysis includes socioeconomic and demographic control variables, using the Index of Multiple Deprivation (IMD) score from the Ministry of Housing, Communities and Local Government\footnote{\url{https://www.gov.uk/government/statistics/english-indices-of-deprivation-2019}} that measures socioeconomic disadvantage across seven dimensions; income, employment, education, health, crime, housing, and living environment. As these scores are only updated around every four years and remain relatively stable over short periods, we apply 2019 values across the analytical timeframe. 

Demographic control variables, derived from ONS annual population estimates and the Annual Population Survey\footnote{\url{https://data.london.gov.uk/dataset/ethnic-groups-borough}}, are selected systematically through both theoretical justification from the public health and criminology literature and empirical validation. The initial pool includes demographic borough-level characteristics covering age structure, gender composition, and ethnic diversity. We calculate Variance Inflation Factors (VIFs) for all potential control variables,
\begin{equation}
    \mathrm{VIF}_i=\frac{1}{1-R_i^2},
\end{equation}
with $R_i^2$ as the coefficient of determination for regressing the $i$-th independent variable on all other independent variables \citep{sheather2009}. We exclude control variables at $\mathrm{VIF} > 5$ to avoid multicollinearity and add the remainder to the model using a stepwise approach, monitoring changes in $R^2$ and coefficient stability. Two demographic control variables are retained; youth demographics as the percentage of population under 17 years and ethnic composition as the population percentage of mixed or other ethnic groups.

Our cross-sectional regression and clustering analyses focus on April 2019 to March 2020 as the most recent financial year with comprehensive data availability, with the annual cycle chosen in alignment with the mental health data reporting structure. Predictive models include the full April 2017 to March 2020 period, with 2017--2019 and 2019-2020 serving as training and test data, respectively. Prior to 2017, differences in mental health data structure restrict cross-year comparability, and a large-scale administrative reorganisation after March 2020 limits subsequent borough-level mental health data availability. Additionally, the onset of the COVID-19 pandemic considerably affected service delivery patterns and data quality.

\subsection{Exploratory Analysis}
\label{subsec:4.3}

Table~\ref{tab:table_1} shows summary statistics for our 2019 data, revealing substantial variation across variables. Overall and property crime rates exhibit particularly high dispersion and highly positive skewness, indicating right-tailed distributions, with several boroughs experiencing disproportionately high crime levels.

\begin{table*}[!h]
\caption{Descriptive statistics for crime rates, mental health referrals, and socioeconomic indicators through IMD scores across 32 London boroughs for 2019 data. Crime and referral rates are expressed per 100,000 population. IMD scores are on a 0-100 scale, with greater deprivation for higher scores.}
\begin{center}
\begin{tabular}{|l|r|r|r|r|r|}
\hline
\textbf{Statistic} & \textbf{Overall Crime} & \textbf{Violent Crime} & \textbf{Property Crime} & \textbf{Referrals} & \textbf{IMD} \\
\hline
\hline
Mean & $6840.024$ & $2874.035$ & $3965.988$ & $3147.194$ & $21.500$ \\
\hline
Median & $6447.445$ & $2887.848$ & $3447.112$ & $2978.899$ & $21.898$ \\
\hline
Std. Dev. & $2350.39$ & $732.197$ & $1689.992$ & $1028.208$ & $6.053$ \\
\hline
Min & $4748.31$ & $1760.313$ & $2743.143$ & $1522.636$ & $9.425$ \\
\hline
Max & $17572.947$ & $5721.613$ & $11851.333$ & $6366.372$ & $32.768$ \\
\hline
Skewness & $3.269$ & $1.739$ & $3.593$ & $1.019$ & $-0.045$ \\
\hline
\end{tabular}
\end{center}
\label{tab:table_1}
\end{table*}

\begin{figure*}[!htb]
    \centering
    \includegraphics[width=\textwidth]{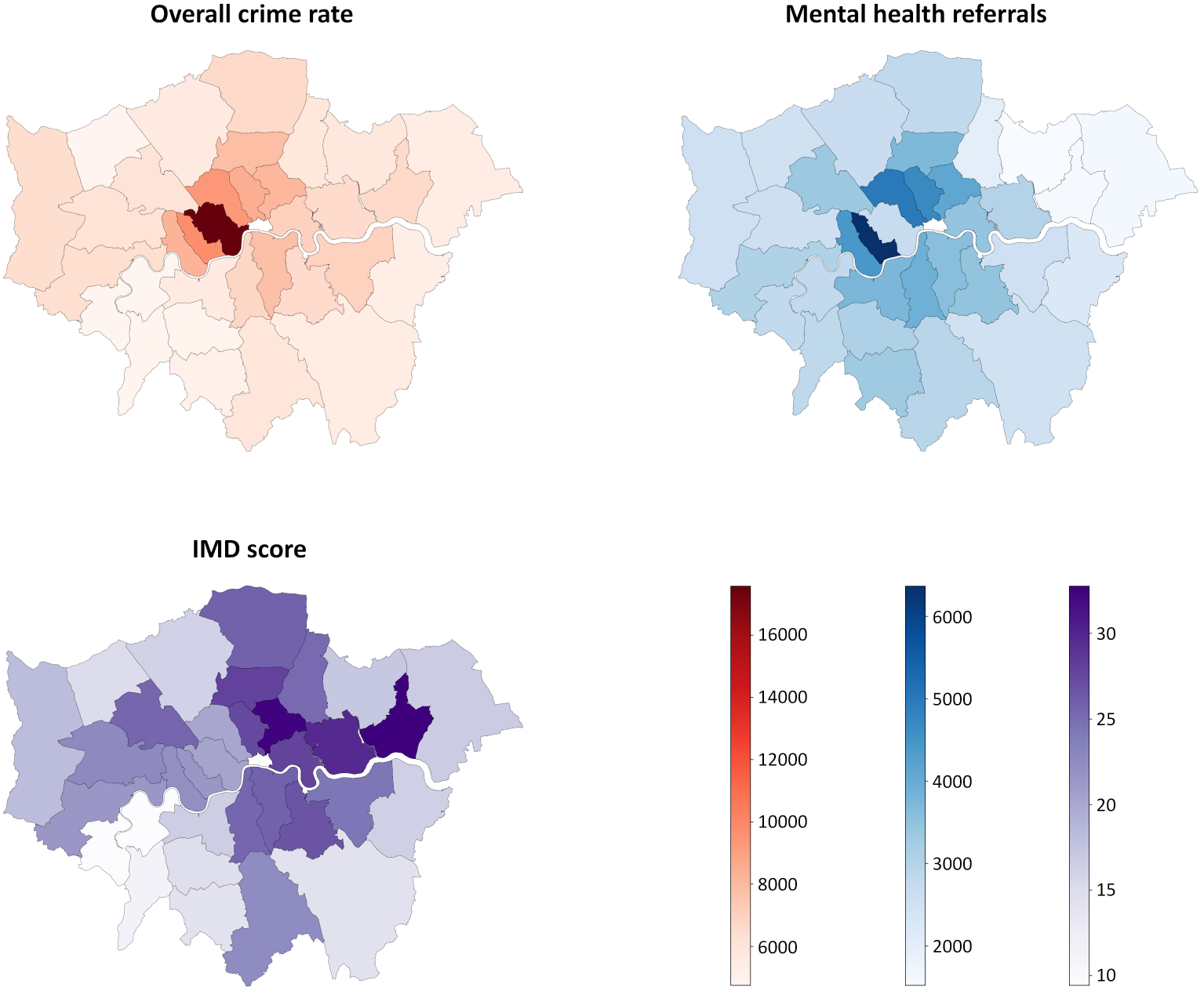}
    \caption{Spatial distribution of variables of interest across London boroughs, excluding the City of London, for 2019. The figures shows crime rates and mental health referrals per 100,000 population on the upper left (red) and upper right (blue), respectively, as well as IMD scores in the lower left (purple).}
    \label{fig:figure_1}
\end{figure*}

This variation is reflected in the choropleth maps of Figure~\ref{fig:figure_1}, with the highest levels in central London, where Westminster and surrounding boroughs form a clear hotspot. Conversely, outer London boroughs, particularly in the southwest and southeast, maintain the lowest crime levels. Violent crime rates follow a similar clustering with an even starker contrast between central and outer London, while property crime rates, still most concentrated in central London, show a more even distribution.

Mental health referral rates show a similar distribution, with the highest rates concentrated in inner London, particularly Kensington and Chelsea. Outer London, especially in the east, exhibits lower rates, suggesting potential gaps, while deprivation via IMD scores shows considerable overlap with crime, with the highest deprivation in north and north-east London. Comparing the geographic patterns of crime and mental health referrals reveals complex relationships, suggesting positive spatial associations rather than the expected inverse; Some moderate-crime areas exhibit high referral rates, while some high-crime areas do not. Rather than indicating simple preventive effects, this indicates shared underlying factors. 

\subsection{Cross-Sectional Analysis}
\label{subsec:4.4}

\begin{figure*}[!htb]
    \centering
    \includegraphics[width=\textwidth]{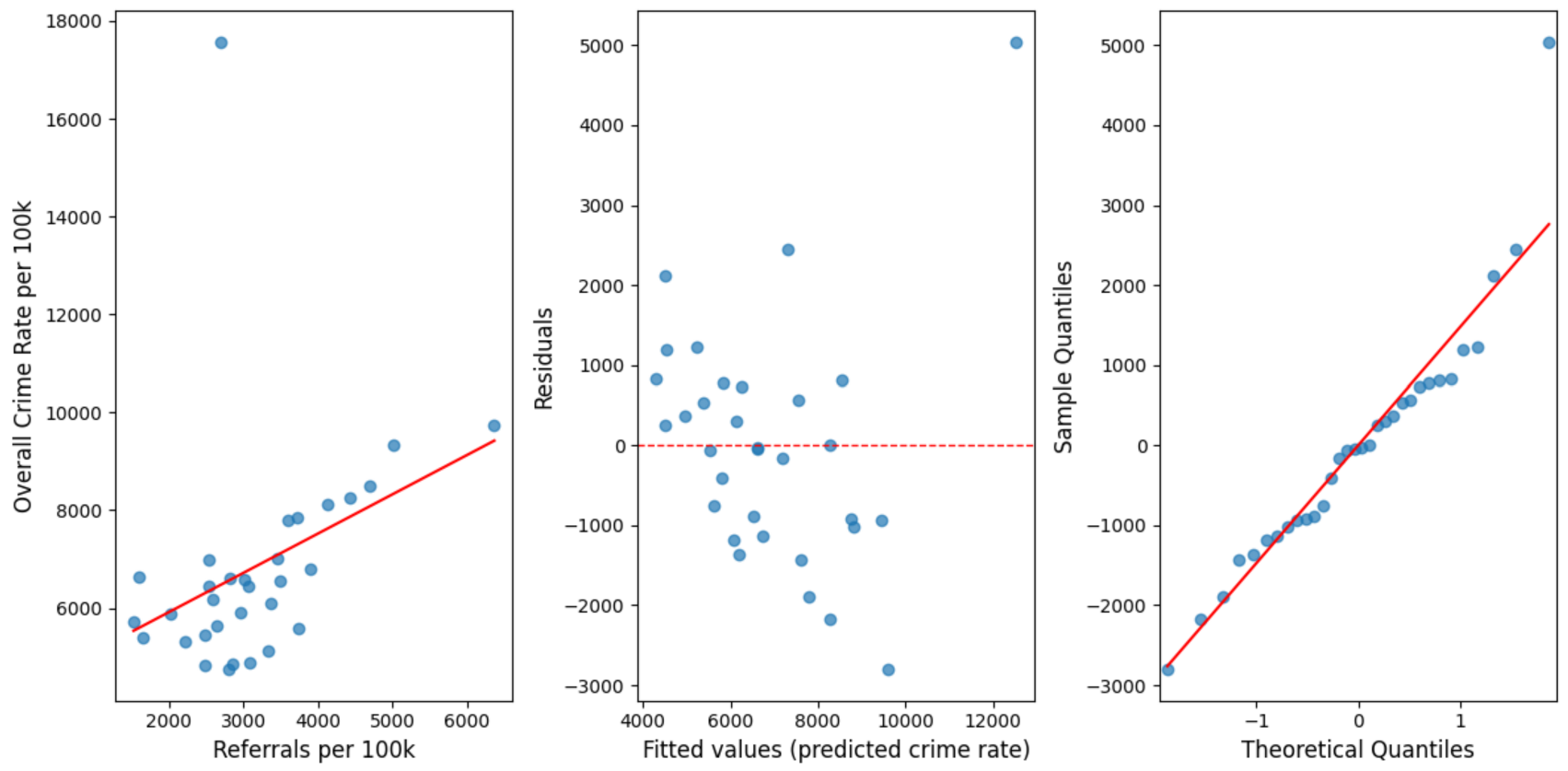}
    \caption{Simple linear regression results for mental health referrals versus crime rate, per 100,000 population, for 2019 data from Greater London. From left to right, the plots show the scatter plot and regression line for the variables, the corresponding residual plot to assess the linearity assumption, and the quantile-quantile (Q-Q) plot for the linear regression to test the normality assumption.}
    \label{fig:figure_2}
\end{figure*}

We initially use a multiple linear regression model for a bivariate analysis of mental health referrals and crime rates, leading to a statistically significant positive relationship with a coefficient of 0.803 ($p = 0.049$). However, the $R^2$ value indicates weak explanatory power, with mental health referrals explaining only 12.3\% of the variance in crime rates. The line fit, as well as residual and Q-Q plots, are shown in Figure~\ref{fig:figure_2}.

When control variables are included, the coefficient for the referral rate changes its sign ($\beta_1$ = -0.819), implying a negative relationship with crime rates, but remaining statistically insignificant ($p = 0.069$) while raising the explanatory power, with an $R^2$ value of 0.595 (see Table~\ref{tab:table_2}). To ensure reliability, we test key assumptions and reveal important violations affecting the interpretation.

\begin{table*}[!h]
\caption{Linear regression results for crime rates and mental health referrals per 100,000 population in 2019, controlling for deprivation (IMD), youth demographics (\% under 17), and ethnic composition (\% mixed/other). Crime rates are expressed per 100,000
population, and standard errors are in parentheses. *** $:= p < 0.05$; ** $:= p < 0.05$; * $:= p < 0.10$.}
\begin{center}
\begin{tabular}{|l|r|r|r|r|}
\hline
& \rule{0pt}{4.5ex}\textbf{Overall Crime} 
& \rule{0pt}{4.5ex}\textbf{\shortstack{Overall Crime\\with Controls}} 
& \rule{0pt}{4.5ex}\textbf{Violent Crime} 
& \rule{0pt}{4.5ex}\textbf{Property Crime} \\
\hline
\hline
referrals\_rate\_per100k & **$0.803$ & *$-0.819$ & **$-0.267$ & *$-0.652$ \\
 & ($0.391$) & ($0.432$) & ($0.120$) & ($0.320$) \\
\hline
imd\_score &  & ***144.322 & ***79.907 & $64.414$ \\
 &  & ($52.375$) & ($14.561$) & ($38.743$) \\
\hline
age\_under\_17\_pct &  & ***$-538.662$ & ***$-147.126$ & ***$-391.537$ \\
 &  & ($167.639$) & ($46.605$) & ($124.005$) \\
\hline
mixed\_other\_pct &  & **$186.409$ & **$46.738$ & **$139.670$ \\
 &  & ($80.931$) & ($22.499$) & ($59.866$) \\
\hline
constant & ***$4313.553$ & ***$1.613e+04$ & ***$4718.797$ & ***$1.141$e+$04$ \\
 & ($1291.833$) & ($4669.479$) & ($1298.142$) & ($3454.082$) \\
\hline
\textit{$R^2$} & $0.123$ & $0.595$ & $0.678$ & $0.572$ \\
\hline
\end{tabular}
\end{center}
\label{tab:table_2}
\end{table*}

The residual plot shows a negative trend, suggesting a more complex relationship, with the the Q-Q plot demonstrating a deviation from normality, confirmed by a Shapiro-Wilk test ($p = 0.045$). Testing for homoscedasticity, Levene's test indicates equal variance of residuals ($p = 0.370$), while a Moran's $I$ of 0.182 ($p < 0.001$) reveals moderate spatial autocorrelation, showing that neighbouring boroughs tend to have similar residual values and violating the residual independence assumption. The results should, therefore, be interpreted cautiously as standard errors and $p$-values may be unreliable, although robustness checks for temporal stability using 2018 data yield very similar results ($\beta_1 = -0.609$, $p = 0.152$). 

Correlation analyses, listed in Table~\ref{tab:correlation_analysis_2019}, highlight a moderately positive linear relationship between mental health referrals and overall crime rates for Pearson's $r$, but stronger correlations observed with Spearman's $\rho$ and Kendall's $\tau$ point towards monotonic non-linearity with robust rank-based correlations. The latter in particular are underestimated by Pearson's $r$ due to linearity and outlier sensitivity. 

\begin{table*}[!h]
\caption{Correlation measures to assess the relationship between mental health referrals and crime rates for 2019. $p$-values are in parentheses. ** = p-value $<$ 0.05, * = p-value $<$ 0.1}
\begin{center}
\begin{tabular}{|l|r|r|r|}
\hline
 & \textbf{Overall Crime} & \textbf{Violent Crime} & \textbf{Property Crime} \\
\hline
\hline
Pearson's $r$ & **$0.351$ & *$0.308$ & **$0.355$ \\
 & (0.049) & (0.087) & (0.046) \\
\hline
Spearman's rho & ***$0.548$ & ***$0.484$ & ***$0.512$ \\
 & $(0.001)$ & $(0.005)$ & $(0.003)$ \\
\hline
Kendall's tau & ***$0.423$ & ***$0.367$ & ***$0.387$ \\
 & $(0.001)$ & $(0.003)$ & $(0.002)$ \\
\hline
\end{tabular}
\end{center}
\label{tab:correlation_analysis_2019}
\end{table*}

Similar patterns exist for violent and property crime, confirmed by a robustness check with 2018 data. We repeat this with first treatment rates instead of referrals, with the same non-linearity with weaker associations and the same non-significance ($\beta_1$ = -1.237, \textit{p} = 0.127), supporting the choice of proxy. Disaggregated linear regression results in Table \ref{tab:table_2} show a higher explanatory power for violent crime, paired with a significant association ($\beta_1 = -0.267$, $p = 0.035$), compared to property crime ($\beta_1 = -0.551$, $p = 0.096$).

Non-linearity has important theoretical and policy implications, as effects may not be constant across accessibility profiles and involve threshold effects or diminishing returns. Positive coefficients appear to contradict the prevention hypothesis, but might instead reveal a demand-driven dynamic rather than supply-driven prevention effects, highlighting service access patterns and a bidirectional relationship.

Testing a polynomial regression model does, indeed, provide a better fit ($R^2 = 0.662$) with statistically significant negative linear and positive quadratic coefficients, indicating a U-shaped relationship. This suggests that at lower access levels, increases in referrals are associated with crime reductions, supporting the prevention hypothesis, but beyond the turning point, increases correspond to rising crime rates, reflecting a demand-driven response for high-crime boroughs. Separate analyses for crime categories show violent crime coefficients to lack statistical significance, and the quadratic specification fails to adequately capture the relationship. Robustness analyses with 2018 data and first treatment rates as an alternative measure result in slightly weaker but consistent patterns, maintaining the U-shaped relationship and level of significance.

\subsection{Predictive Modelling with Decision Trees}
\label{subsec:4.5}

When using the polynomial regression model, trained on 2017--2018 data, for predictions based on 2019 data, the model achieves an average prediction error of ${\sim}15.4\%$ ($R^2 = 0.573$, $\mathrm{MAE} = 1,034.12$), but a five-fold CV reveals severe overfitting, with a mean $R^2$ of $-0.188$, underperforming simple mean prediction. Likely due to the small sample size combined with the presence of outliers and the complexity introduced by the quadratic terms, this result holds for a robustness check using first treatment rates. 

\begin{figure*}[!htb]
    \centering
    \includegraphics[width=\textwidth]{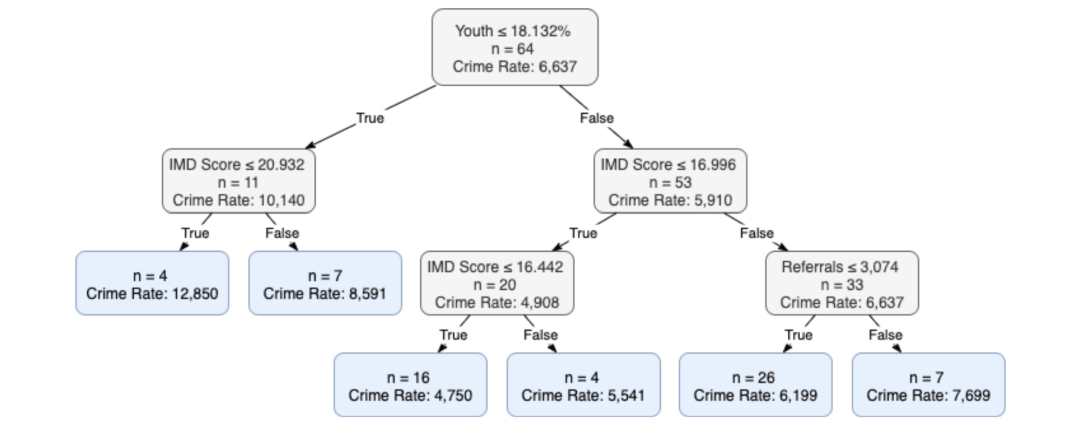}
    \caption{Decision tree for overall crime rate prediction for Greater London data. The figure shows hierarchical splitting rules based on youth demographics, deprivation (IMD), and mental health referrals. Terminal nodes display predicted crime rates and sample sizes.}
    \label{fig:figure_3}
\end{figure*}

An initial unconstrained model decision tree model leads to a structure with $127$ nodes, a tree depth of $12$, and $64$ leaves, outperforming the polynomial regression model ($R^2 = 0.815$, $\mathrm{MAE} = 659.854$) but lacking interpretability \citep{rudin2019}. We implement a grid search with five-fold cross-validation, resulting in a maximum tree depth of $3$, minimum samples per split of $2$, and minimum samples per leaf of $4$. The resulting model, shown in Figure~\ref{fig:figure_3}, achieves an average prediction error of ${\sim}10.4\%$ ($R^2 = 0.723$, $\mathrm{MAE} = 699.725$), maintaining a higher performance than the polynomial model.

The tree's root node sets youth demographics as the primary predictor. While consistent with literature emphasising young people's role in crime demographics \citep[see, for example,][]{farrington2003}, boroughs with lower youth proportions and lower deprivation levels feature the highest predicted crime rates, which may reflect unique characteristics such as commercial activity and tourism. In contrast, boroughs with similar youth demographics but higher deprivation levels show lower (but still elevated) predicted crime.

For the other side, the model uses more complex decision pathways involving both deprivation and mental health service access. Crime rates for areas with very low deprivation are predicted to be moderate, whereas higher youth populations and moderate levels of deprivation lead to a distinction between lower and higher mental health referral rates. If the latter is lower, predicted crime rates are lower as well, and vice versa. While this appears inconsistent with the prevention hypothesis, it might reflect a more reactive demand-driven increase in service access consistent with previous findings.

\begin{figure*}[!htb]
    \centering
    \includegraphics[width=\textwidth]{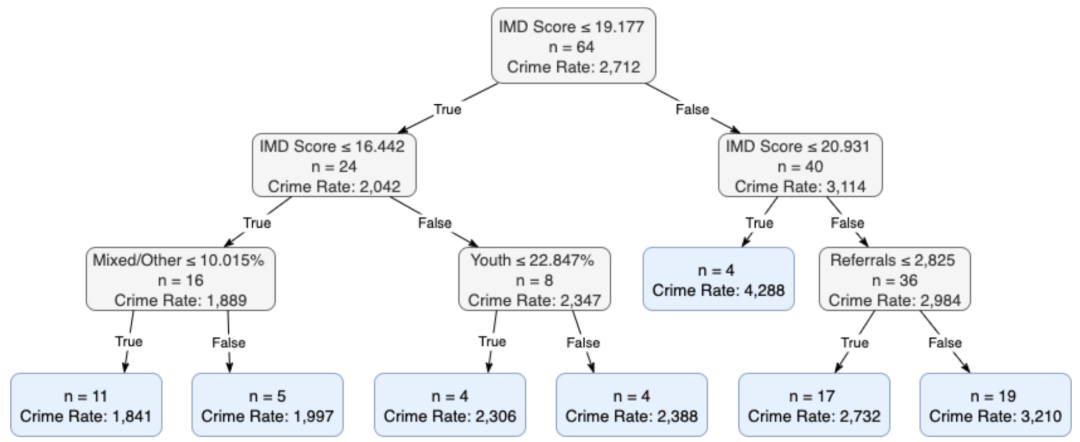}
    \caption{Decision tree for violent crime rate prediction for Greater London data. The figure shows hierarchical splitting rules led by deprivation (IMD), with further splitting based on ethnic composition and mental health referrals. Terminal nodes display predicted crime rates and sample sizes.}
    \label{fig:figure_4}
\end{figure*}

Overall, the model structure suggests a particular relevance of mental health service access in areas with specific combinations of youth demographic and deprivation levels. We then examine whether the predictive patterns hold across crime categories, building separate models for violent and property crime. The former yields an average prediction error of ${\sim}9.7\%$ ($R^2 = 0.700$, $\mathrm{MAE} = 267.079$), with a markedly different hierarchical structure shown in Figure~\ref{fig:figure_4}. 

Both youth population and deprivation continue to play a key role, but with different order and splitting thresholds, placing deprivation levels at the root node. Notably, the tree only introduces these for the least deprived areas, which could reflect differentiating effects of racial discrimination in affluent areas, whereas youth demographics and mental health referral rates otherwise remain the splitting criteria.

\begin{figure*}[!htb]
    \centering
    \includegraphics[width=\textwidth]{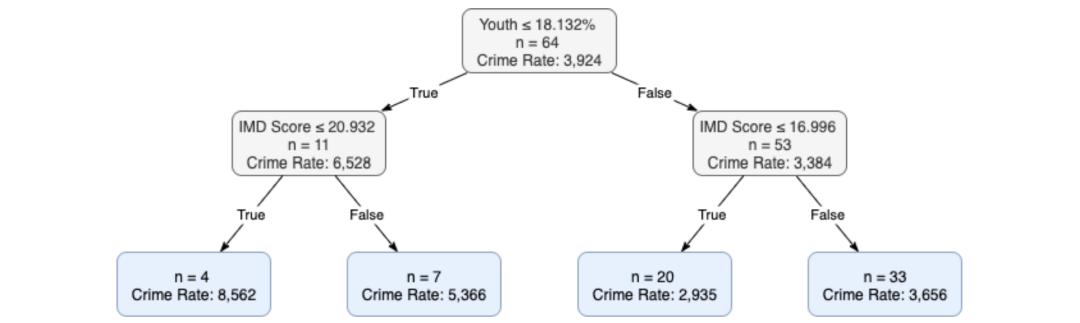}
    \caption{Decision tree for property crime rate prediction for Greater London data. The figure shows hierarchical splitting rules based on youth demographics and deprivation (IMD). Terminal nodes display predicted crime rates and sample sizes.}
    \label{fig:figure_5}
\end{figure*}

The property crime model, shown in Figure~\ref{fig:figure_5}, performs similarly, with an average prediction error of ${\sim}13.5\%$ ($R^2 = 0.714$, $\mathrm{MAE} = 532.768$). While the accuracy is slightly lower, the root node retains youth demographics at the root node. The structure, however, is notably simpler at only two levels, suggesting that the predictable component of property crimes is based on a more streamlined set of factors.

We perform a robustness test for overall crime rates, using 2016--2017 data to predict 2018 rates, with an average prediction error of ${\sim}12.0\%$ ($R^2 = 0.579$, $\mathrm{MAE} = 806.97$). The resulting structure is identical, suggesting that predictive relationships are temporally robust. Additionally, a robustness check for first treatment rates as an alternative measure also leads to the same decision tree structure, with an average prediction error of ${\sim}11.0\%$ ($R^2 = 0.710$, $\mathrm{MAE} = 737.42$), further solidifying this finding. 

\subsection{Cluster Analysis for Regional Typologies}
\label{subsec:4.6}

\begin{figure*}[!htb]
    \centering
    \includegraphics[width=\textwidth]{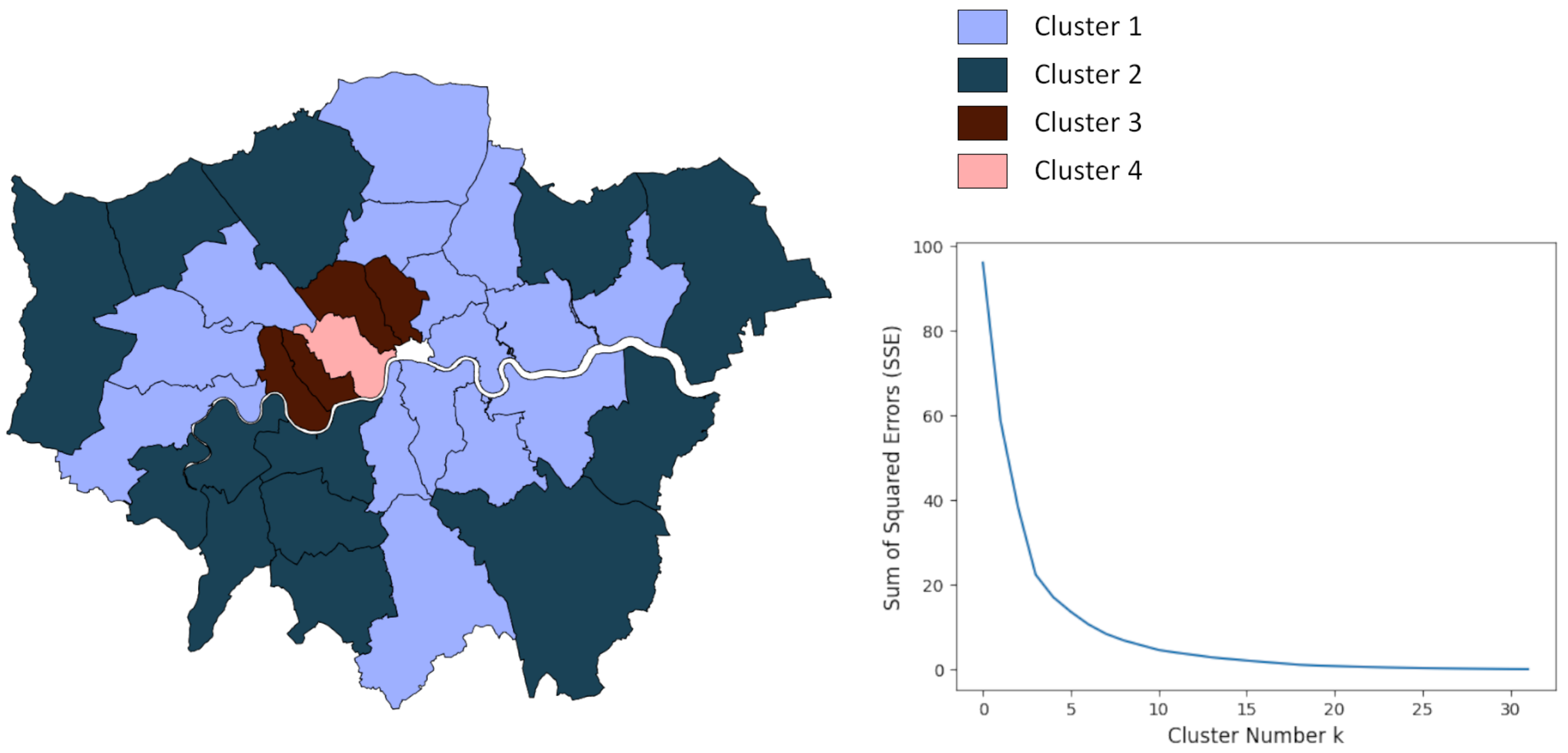}
    \caption{Cluster analysis results for Greater London, using crime rates, mental health referral rates, and deprivation (IMD). The left panel shows the demographic distributions of clusters, with a legend in the upper right, whereas the lower right panel depicts the elbow criterion as the number of clusters versus the within-cluster sum of squared errors.}
    \label{fig:figure_6}
\end{figure*}

As the previous model-building implicitly assumes uniform relationships, we identify borough typologies with similar patterns using $k$-means clustering as a follow-up to the moderate degree of spatial clustering identified through Moran's $I$. We use both the elbow criterion and average silhouette scores ($k = 3: 0.257$, $k = 4: 0.292$, $k = 5: 0.216$, $k = 6: 0.216$) for an optimal number of four clusters shown in Figure~\ref{fig:figure_6}, balancing explanatory power and interpretability. Using the Kruskal-Wallis test, all three key variables are identified as statistically significant, indicating that the clusters reflect meaningful variation in the underlying data. Separate analyses of violent and property crime rates lead to identical cluster assignments.

Typologies are summarised in Table~\ref{tab:cluster_characteristics}, with Cluster 1 as the largest group of 15 boroughs across inner and outer London and characterised by moderate crime and referral rates, as well as the highest levels of deprivation. These include Barking and Dagenham, Brent, Croydon, Ealing, Enfield, Greenwich, Hackney, Haringey, Hounslow, Lambeth, Lewisham, Newham, Southwark, Tower Hamlets, and Waltham Forest.

\begin{table*}[!h]
\caption{Summary of identified clusters in Greater London, showing the number of boroughs, crime rates for violent, property, and overall crime rates (per 100,000 population), mental health referral rates (MHR), and deprivation scores (IMD). IMD scores range 0--100, with higher values indicating greater deprivation.}
\begin{center}
\begin{tabular}{|l|r|r|r|r|r|r|}
\hline
\textbf{Cluster} & \textbf{Boroughs} & \textbf{Violent Crime} & \textbf{Property Crime} & \textbf{Overall Crime} & \textbf{MHR} & \textbf{IMD} \\
\hline
\hline
1 & 15 & 3,086.1 & 3,676.5 & 6,762.5 & 3,080.5 & 26.42 \\
\hline
2 & 12 & 2,216.3 & 3,118.6 & 5,334.9 & 2,610.9 & 14.99 \\
\hline
3 & 4 & 3,340.2 & 5,622.5 & 8,962.7 & 5,121.5 & 22.87 \\
\hline
4 & 1 & 5,721.6 & 11,851.3 & 17,572.9 & 2,686.5 & 20.34 \\
\hline
\end{tabular}
\end{center}
\label{tab:cluster_characteristics}
\end{table*}

Cluster 2 contains 12 boroughs primarily located in outer London, with the lowest crime and referral rates as well as deprivation levels, including Barnet, Bexley, Bromley, Harrow, Havering, Hillingdon, Kingston upon Thames, Merton, Redbridge, Richmond upon Thames, Sutton, and Wandsworth. Cluster 3 contains four boroughs in Central London, with high crime and referral rates and moderate deprivation levels, covering Camden, Hammersmith and Fulham, Islington, and Kensington and Chelsea. 

Lastly, Cluster 4 consists only of Westminster, with extremely high crime rates substantially exceeding other clusters, showing moderate referral rates and deprivation levels. This area reflects the unique characteristics of Westminster as one of London's main commercial, governmental, and tourist districts, reflecting very similar findings for the Chicago Loop district \citep[see, for example,][]{moews2021}.

Examining additional variables, the Kruskal-Wallis test reveals significant differences across clusters, with the most pronounced contrast observed in the youth population. Clusters 1 and 2 exhibit substantially higher youth demographics, suggesting an effect on regional patterns of mental health service access and crime rates. Robustness checks confirm temporal cluster stability, with 2018 data leading to identical cluster assignments, but using first treatment rates shows a difference in assignments, indicating that, while reasonably interchangeable in previous experiments, both proxies capture distinct spatial patterns.


\section{Discussion}
\label{sec:discussion}

Existing literature on mental health service access and crime rates predominantly estimates the size and direction of relationships using traditional statistical methods that assume additive and constant effects across contexts \citep{deza2022b}. While the criminology literature has increasingly emphasised the exploration of functional form, they are largely centred on associations between crime and environmental or socioeconomic factors \citep{kim2023, walker2007, hipp2011}. 

This work adds to this growing body of research by studying mental health service access and crime rates in the Greater London area, simultaneously contributing to the expansion of a literature with a dominant focus on the United States. Our experiments consistently point towards complex non-linear and consistently positive relationships, contradicting the prevention hypothesis underlying these studies \citep{wagner2020}. They reveal a more nuanced U-shaped relationship, suggesting threshold-dependent effects with negative associations below the threshold, consistent with the literature, while positive relationships above the threshold could reflect a demand-driven rather than supply-driven dynamic.

Crime can have long-lasting impacts, requiring continuous support extending beyond direct victims, with community-level impacts on mental health and fear of crime that similarly increase service demand \citep{russo2010, fowler2009, jackson2009, chandola2001}. Mental health issues have also been found to increase victimisation risk, thus reinforcing the demand cycle \citep{choe2008}.

This demand-driven interpretation points to an important limitation, as referrals as a proxy for service access may include this factor, which is also reflected in the clustering of distinct borough typologies. Except for Westminster, clusters show corresponding relative levels of crime rates and mental health referrals, where high-crime areas align with high referral rates (Clusters 1 and 3), and vice versa (Cluster 2). The results, therefore, require careful interpretation but offer valuable insights for policy development. 

Disaggregation into violent and property crimes reveals distinct patterns consistent with the criminology literature, where population-level prevention studies show stronger prevention effects for property crimes \citep{wagner2020, hendrix2022}. Empirical studies generally ascribe more rational decision-making, making them more susceptible to policy changes and interventions, whereas violent crime tends to be more impulsive and affected by situational factors \citep{chalfin2019, cornish2016}.

This is reflected in both in the initial polynomial regression, with statistically significant coefficients only for property crime, and the primary experiments using decision trees. For property crime, the predictive structure is simpler, involving only youth demographics and deprivation, while that for violent crime shows greater complexity, with multiple levels of demographic, socioeconomic, and service access variables. 

Our experiments identify deprivation as the primary predictor, in line with the public health literature \citep{kelly2000, sampson1995}. These differential patterns suggest the need for tailored approaches, with property crime interventions focussing on rational choice-based strategies reducing opportunities, while violent crime likely requires comprehensive structural interventions addressing deprivation, especially given their substantially higher social costs \citep{miller1996}.

IMD scores demonstrate a positive relationship with overall and violent crime rates, but our analysis shows no statistically significant association with property crimes, seemingly contradicting existing research \citep[see, for example,][]{allen1996}. This could reflect two opposing effects of economic deprivation on property crime as described by crime opportunity theory; while deprivation causes strain and disorganisation to increase property crimes, it simultaneously reduces the number of targets in that area \citep{hannon2002}. In London, this might explain why property crime rates are higher in affluent central areas with more opportunities, driven by tourism, transient population, and commercial density.

While existing research specifies that young people are more likely to engage in criminal activity \citep[see][]{sweeten2013}, our results maintain the predictive power of that variable but persistently link higher youth populations to lower crime rates, highlighting an important limitation. The literature identifies 15–-19-year-olds as the highest-risk demographic \citep[see, for example,][]{farrington2003}, but our data's age bracket includes residents below the age of 17, likely averaging high-risk adolescents with low-risk younger children. Future research as well as policy work should address this limitation of data availability.

The analysis also demonstrates a positive association between mixed and other ethnicity proportions and crime rates, aligning with extensive research on economic and social disadvantages of ethnic minority communities affecting neighbourhood crime and violence \citep{pratt2005}. Structural disadvantage is a key factor in explaining variations in crime rates, as minority populations are disproportionately exposed to criminogenic conditions including residential segregation, limited economic opportunities, and high levels of poverty \citep{peterson2010, sampson1995}. 

Cluster analyses provides evidence for this structural interpretation in London, with Cluster 1 experiencing the highest deprivation levels and also features higher proportions of residents in that category, while Cluster 2, with the lowest deprivation, has the lowest proportions of these groups. These patterns indicate that ethnic composition reflects broader structural inequalities.

The findings discussed above offer further insights for interventions. The strong association between deprivation and crime, particularly violent crime, supports structural interventions, addressing underlying socioeconomic disadvantages. Similarly, as the patterns observed in ethnically diverse areas reflect broader structural inequalities, these require interventions that reduce the socioeconomic disadvantage for ethnic minorities. However, for age-targeted interventions, no reliable policy implications can be derived due to the measurement limitation of the youth population variable.

Given the above discussion, policy recommendations for Cluster 1, characterised by the highest deprivation and relatively high violent crime levels, include addressing underlying factors through socioeconomic interventions such as increased social housing as well as employment and education opportunities. In contrast, Cluster 2 represents low-crime and low-referral borough, where policies should maintain existing service levels while monitoring to identify crime rate changes. Detailed analyses of these boroughs could help to identify the most successful prevention approaches to develop strategies for others. 

Cluster 3 includes four Central London boroughs with elevated transient population levels, tourism, and commercial activity, which is reflected in higher property crime rates. Here, policy-making could focus on addressing crime opportunities through improved environmental design, enhanced security measures, and crime prevention education. Simultaneously, these boroughs have the highest referral rates, which likely reflects demand-driven patterns due to exposure. While moderately high deprivation levels indicate underlying vulnerabilities, these could be enhanced by the unique urban environment, including high residential density and housing instability. Therefore, interventions should include broad community programmes addressing these urban stressors while improving service integration and effectiveness to create comprehensive support systems, addressing both the elevated crime rates and community well-being.

Cluster 4, or Westminster, as one of the main Central London boroughs, exhibits considerable transient population and tourism levels accompanied by disproportionately high crime rates. With a large proportion being property crimes, policy should prioritise opportunity reduction as for Cluster 3, but the borough likely requires more intensive interventions, including coordination with transport police and tourism authorities, as community-based interventions may be less effective.

Overall, machine learning techniques can provide crucial insights into the relationship between mental health service access and crime rates, with a growing number of researchers applying these models to capture complex non-linear dynamics, interactions, and threshold effects \citep{kim2023, lee2024, zhang2022}. While the eschewal of black-box models proves useful, some limitations constrain the generalisability of our findings. The issue of data availability requires the use of a proxy variable, which future research can potentially address by developing alternative mental health access measures, for example following \citet{deza2022b} regarding the number of office-based services, although this introduces assumptions about residence and service locations. 

These experiments should also be repeated once a sufficient data timeframe of post-pandemic data is available to explore altered patterns of both mental health and crime \citep{hoeboer2023, byrne2021}. Finally, the unique characteristics of the Greater London area may limit generalisability to sufficiently different urban contexts as well as international settings; an expansion of these techniques to other areas could further explore the investigated relationship.


\section{Conclusion}
\label{sec:conclusion}


This study examines the relationship between access to mental health services and crime rates in the Greater London area, addressing a critical gap in understanding the effect of mental health interventions within a universal healthcare system. Using traditional statistical methods and interpretable machine learning techniques, our analysis uncovers complex relationships across 32 boroughs. The results challenge prevailing assumptions; instead of an expected negative relationship, with increased service access corresponding to reduced crime, we identify positive associations across different analytical approaches.

These patterns suggest that our mental health proxy includes service demand in response to crime exposure, with a polynomial regression model indicating a more nuanced U-shaped relationship. While preventive effects may exist at lower service levels, supporting the crime prevention hypothesis, higher referral rates could reflect demand-driven responses, leading to a more complex picture than established by previous US-based studies. Given current data limitations, these relationships warrant further research for the UK setting, using measures that focus on service accessibility instead of demand.

Our prediction of property crime results in simpler decision pathways, likely reflecting a more rational and opportunity-driven nature, and statistically significant associations with mental health service demand are limited to this crime category. Violent crime exhibits greater complexity involving multiple demographic and socioeconomic variables, showing a need for distinct prevention strategies for different categories.

We find that deprivation is a strong predictor of crime rates, especially for violent crime, but broader structural inequalities and socioeconomic disadvantages are also reflected in a less pronounced association with further demographic factors, highlighting the importance of addressing underlying social determinants and inequalities. An investigation of regional typologies through a cluster analysis results in four distinct borough types, which are characterised by unique combinations of mental health service access, crime rates, and deprivation levels, providing empirical evidence of spatial heterogeneity. 

Our research makes several contributions to the understanding of crime prevention within a universal healthcare system. By revealing complex and context-dependent relationships between mental health service access and crime, we demonstrate that effective urban crime prevention requires data-driven and locally adapted strategies, with interpretable machine learning techniques as a useful tool to uncover spatial patterns for informed policy-making. The findings indicate that, within London and the context of the NHS, the investigated relationship might operate through different mechanisms than previously assumed, highlighting the need to re-examine mental health interventions and public safety.

\begin{footnotesize}

\bibliographystyle{apalike}
\bibliography{references}

\end{footnotesize}

\end{document}